\documentclass[aps,prc,showpacs,superscriptaddress,amsmath,amssymb,twocolumn,floatfix]{revtex4}

\usepackage{graphicx}
\usepackage{amsmath}
\usepackage{amssymb}
\usepackage{xcolor}
\usepackage[normalem]{ulem}
\usepackage{multirow}
\usepackage{etoolbox}
\usepackage[colorlinks=true,linkcolor=blue,citecolor=blue,urlcolor=blue]{hyperref}

\makeatletter
\AtBeginDocument{%
  \renewcommand{\NAT@open}{\textcolor{blue}{[}}%
  \renewcommand{\NAT@close}{\textcolor{blue}{]}}%
  \renewcommand{\NAT@sep}{\textcolor{blue}{,}}%
}
\makeatother

\newcommand{\zhoubegin}[1]{\begingroup\color{blue}}

\newcommand{\Hg}{\ensuremath{^{180}\mathrm{Hg}}}
\newcommand{\Hgs}{\ensuremath{^{180}\mathrm{Hg}^{\ast}}}

\newcommand{\Ar}{\ensuremath{^{36}\mathrm{Ar}}}

\newcommand{\Zr}{\ensuremath{^{90}\mathrm{Zr}}}

\newcommand{\bs}{\boldsymbol}

\definecolor{V09Change}{RGB}{0,105,86}

\begin{document}

\title{Role of the scission-line structure in the formation of the
fragment-mass distribution in $^{180}$Hg fission}

\author{Jin-Tian Liu}
\affiliation{Institute of Theoretical Physics, Chinese Academy of Sciences, Beijing 100190, China}
\affiliation{School of Physical Sciences, University of Chinese Academy of Sciences, Beijing 100049, China}

\author{A. Rahmatinejad}
\affiliation{Joint Institute for Nuclear Research, Dubna, 141980, Russia}

\author{Jie Zhao}
\affiliation{Engineering Research Center of Advanced Functional Materials and Intelligent Sensing, 
Education Department of Guangxi Zhuang Autonomous Region, 
and Guangxi Key Laboratory of Nuclear Physics and Technology, 
Guangxi Normal University,Guilin 541004, China}

\author{T.~M.~Shneidman}
\affiliation{Joint Institute for Nuclear Research, Dubna, 141980, Russia}

\author{N. V. Antonenko}
\affiliation{Joint Institute for Nuclear Research, Dubna, 141980, Russia}

\author{Shan-Gui Zhou}\email{sgzhou@itp.ac.cn}
\affiliation{Institute of Theoretical Physics, Chinese Academy of Sciences, Beijing 100190, China}
\affiliation{School of Physical Sciences, University of Chinese Academy of Sciences, Beijing 100049, China}
\affiliation{School of Nuclear Science and Technology, University of Chinese Academy of Sciences, Beijing 100049, China}

\begin{abstract}
The fission of $^{180}\mathrm{Hg}$ is investigated by studying
the stochastic evolution on the potential energy surface calculated within the multidimensionally constrained relativistic Hartree--Bogoliubov
(MDC-RHB) model. The MDC-RHB model provides the
zero-temperature potential-energy surface in the $(\beta_{20},\beta_{30})$
plane and self-consistent density distributions, from which scission
configurations are identified using a geometrical neck criterion.
The evolution toward scission is described by a master equation on the
same deformation mesh, with transition rates determined by
deformation-dependent microscopic level densities.
The level densities are obtained in a finite-temperature
BCS treatment using the MDC-RHB single-particle
spectra and zero-temperature pairing gaps as input. 
The
effective barrier-top region is found to lie close to the scission line,
making the detailed energetic structure of the scission boundary
particularly important for the fragment distribution. The potential-energy
landscape and scission-line structure strongly favor reflection-asymmetric
configurations with pronounced prefragment deformations and lead to
fragment-mass distributions centered close to the experimentally observed
$A_{\mathrm L}/A_{\mathrm H}\simeq80/100$ division. An additional
constrained $\beta_{40}$ analysis indicates that fluctuations beyond the
$(\beta_{20},\beta_{30})$ collective space can generate an additional
spread of the fragment masses. With increasing excitation energy over the
range $E^*=34.0$--$65.5$ MeV, the symmetric component is enhanced, while
the asymmetric fission mode remains dominant.
\end{abstract}

\pacs{
Keywords: fission, \Hg, {\color{black}MDC-RHB model}, finite-temperature BCS, level density}

\maketitle

\section{Introduction}

The $\beta$-delayed fission of $^{180}$Tl, proceeding through excited
states of $^{180}\mathrm{Hg}$, revealed pronounced mass-asymmetric
fission in the neutron-deficient lead region
\cite{Andreyev2010,Elseviers2013,Andreyev2013}.
Before this measurement, the intuitive expectation had been a
mass-symmetric division of the extremely neutron-deficient nucleus
\Hg{} into two identical \Zr{} fragments with the magic neutron number
\(N=50\) and the semimagic proton number \(Z=40\).
The measured fragment distribution instead showed a pronounced
asymmetric split close to
\(A_{\mathrm{L}}/A_{\mathrm{H}}\simeq80/100\),
revealing a mechanism different from the familiar actinide case where
fragment magic numbers around \(Z=50\) and \(N=82\) play the dominant role.
Since then, fission of neutron-deficient nuclei in the lead region has
attracted extensive experimental
\cite{Nishio2015,Kumar2020,Bogachev2021,Kozulin2021,Kozulin2022,Vardaci2024}
and theoretical attention
\cite{Ichikawa2012,Moller2012,Warda2012,McDonnell2014,MollerRandrup2015,
Scamps2019,Li2022,Andreev2012,Andreev2013,Pasca2020}.
In particular, subsequent fusion-fission measurements of \Hgs{} formed
in \(\Ar + ^{144}\mathrm{Sm}\) reactions have confirmed that the asymmetric
mode observed in $\beta$-delayed fission also persists in fusion-induced
fission at moderate excitation energies above the fission barrier
\cite{Nishio2015,Kumar2020,Bogachev2021,Kozulin2022}.

Theoretical studies have attributed this surprising behavior to the
topology of the multidimensional potential-energy surface (PES) and to
shell effects in deformed prescission configurations.
Macroscopic-microscopic calculations by Ichikawa, M\"{o}ller, Randrup,
and collaborators showed that the asymmetric fission valley of \Hg{}
is separated from the symmetric fusion valley by a ridge, and Brownian
shape evolution calculations reproduced asymmetric yields in
neutron-deficient Hg isotopes
\cite{Ichikawa2012,Moller2012,MollerRandrup2015}.
Self-consistent Skyrme and Gogny density functional calculations reached
a similar conclusion: The favored path proceeds through elongated,
reflection-asymmetric configurations, whereas the symmetric
\(\Zr+\Zr\) channel lies in a different valley
\cite{Warda2012,McDonnell2014}.
Scission-point and fragment-based analyses further emphasized that
deformed fragments can be energetically favored over the apparently
attractive spherical symmetric split
\cite{Andreev2012,Panebianco2012,Andreev2013}.
More recently, microscopic calculations have identified deformed shell
gaps and octupole correlations in the nascent fragments as key
ingredients of the asymmetric path \cite{Scamps2019,Li2022}.
In particular, a time-dependent generator-coordinate method calculation
in the Gaussian overlap approximation (TDGCM+GOA), based on covariant
density functional theory (CDFT) with PC-PK1 \cite{Zhao2010}, identified
an asymmetric valley in the $(\beta_2,\beta_3)$ plane.
Additional constrained calculations in $(\beta_2,\beta_3,q_N)$ space
were used to determine scission configurations, with fragments close
to $^{101}\mathrm{Rh}/{}^{79}\mathrm{Br}$.
The study also highlighted the important role of a neutron octupole
shell gap around $N=56$ \cite{Li2022}.
A recent Skyrme density functional study further highlighted the role
of the hexadecapole coordinate in obtaining continuous PESs and
identifying scission configurations in \Hg{} \cite{Su2025Hexadecapole}.

Scission-point models describe the asymmetric mass division of \Hg{}
in terms of the statistical properties of configurations near scission
\cite{Andreev2012,Andreev2013,Pasca2020}.
The scission line, defined as the set of configurations at which the
nucleus is assumed to separate into two fragments, links the collective
deformation landscape to the masses and shapes of the nascent fragments.
Its energetic and statistical structure can therefore favor particular
mass divisions.
In this work, we aim to clarify the role of the scission-line structure
relative to that of the preceding shape evolution in the formation of
the fragment-mass distribution in \Hg{} fission and in its dependence
on excitation energy.

CDFT provides a microscopic framework for describing the evolution
of fissioning nuclei.
It treats nuclei in terms of self-consistent nucleonic densities and
fields within a Lorentz-covariant framework, incorporates the spin-orbit
potential naturally, and has been widely applied to describe
ground-state properties, collective excitations, exotic nuclei,
and fission
\cite{Walecka1974,SerotWalecka1986,Reinhard1989,Ring1996,Vretenar2005,
Meng2006,Paar2007,Niksic2011,Meng2016_RDFNS}.
The multidimensionally constrained covariant density functional theory
(MDC-CDFT) has been developed to study various nuclear shapes and their
manifestations in nuclear structure and fission
\cite{Lu2012,Lu2014}.
The MDC-CDFT framework has been successfully applied to the description
of fission barriers of actinide nuclei \cite{Lu2012,Lu2014,GorielyRIPL4}.
The general MDC-CDFT framework \cite{Zhou2016,Zhao2017} imposes $V_4$ symmetry, which allows
multipole deformations $\beta_{\lambda\mu}$ with even $\mu$ to be included
self-consistently, in particular the triaxial $\beta_{22}$ and
reflection-asymmetric $\beta_{30}$ modes relevant to fission.
The present calculations impose axial symmetry while allowing reflection
symmetry to break.
Depending on the treatment of pairing correlations, two variants have
been developed: the multidimensionally constrained relativistic
mean-field (MDC-RMF) model with the Bardeen-Cooper-Schrieffer (BCS)
approximation \cite{Lu2012,Lu2014} and the multidimensionally constrained
relativistic Hartree-Bogoliubov (MDC-RHB) model with the Bogoliubov
transformation \cite{Zhao2017}.

The MDC-CDFT framework has also been applied to spontaneous and induced fission dynamics
\cite{Zhao2015SpontaneousFission,Zhao2016,Zhao2019FT,Zhao2019,Zhao2020,
Zhao2021,Zhao2022Dissipation,Zhao2022TKE},
third minima and triple-humped barriers in light actinides
\cite{Zhao2015},
shapes and PESs of superheavy nuclei \cite{Meng2020,Wang2022SHN},
nonaxial octupole \(Y_{32}\) and related multipole correlations
\cite{Zhao2012Y32,Zhao2017,Zhao2024ExoticOct,Zhao2024Multipole},
structures of hypernuclei
\cite{Lu2011Hyper,Lu2014Hyper,Rong2020,Chen2021,Sun2022},
low-lying excited states associated with exotic nuclear shapes
\cite{WangLu2022,Rong2023,Lu2025},
\(\alpha\), \(2\alpha\), and heavier-cluster radioactivity
\cite{Mercier2020Alpha,Mercier2021TwoAlpha,Zhao2023ClusterDecay},
and octupole correlations in radon isotopes \cite{Jiang2026}.
The broad range of successful applications demonstrates the capability
of the MDC-CDFT framework to describe multidimensional PESs and competing
shape configurations governed by shell effects.
It is therefore well suited to \Hg{}, where the competition between
symmetric and asymmetric fission paths is closely connected to the
deformation-dependent shell structure.

To address the excitation-energy dependence of the fragment yields,
we combine the deformed single-particle spectra obtained from the
MDC-RHB calculation with a finite-temperature BCS treatment to determine
the local thermodynamic quantities and microscopic level densities at
the intrinsic excitation energy available at each point of the PES.
The resulting deformation-dependent level densities are used to
determine the transition rates between neighboring configurations,
and the shape evolution toward scission is described by a master
equation in the \((\beta_{20},\beta_{30})\) space.
The resulting fragment-mass distributions are compared with those
obtained assuming equilibrium on the same scission line to assess
the influence of the preceding shape evolution.

The paper is organized as follows.
Section~\ref{sec:theoretical_framework} introduces the MDC-RHB model for
calculating the PES, the finite-temperature BCS method for obtaining
intrinsic level densities, and the master equation for describing
shape evolution from the ground-state region to scission.
Section~\ref{sec:results} gives the numerical details and discusses the
potential-energy surface, scission configurations, and excitation-dependent
fragment-mass distributions, including the effect of $\beta_{40}$ variations.
Section~\ref{sec:summary} summarizes the main findings and outlines
future developments.

\section{Theoretical Framework}
\label{sec:theoretical_framework}

\begingroup
\hypersetup{linkcolor=blue,citecolor=blue}
This section presents the MDC-RHB equations, the
finite-temperature BCS expressions for intrinsic level densities, and
the master equation for shape evolution. The interaction parameters,
symmetry restrictions, and numerical implementation are specified in
Sec.~\ref{sec:numerical_details}.
\par\endgroup

\subsection{Multidimensionally constrained relativistic Hartree-Bogoliubov calculation}
\label{sec:mdcrhb}

\begingroup
\color{black}
\hypersetup{linkcolor=blue,citecolor=blue}
Covariant density functional theory (CDFT) describes the nuclear many-body
system in terms of self-consistent scalar and vector fields constrained by
Lorentz symmetry, thereby incorporating relativistic effects and the
spin--orbit interaction within a unified microscopic framework
\cite{SerotWalecka1986,Ring1996,Vretenar2005,Meng2006,Niksic2011,Meng2016_RDFNS}.
The effective interaction can be represented either by finite-range meson
exchange or by local point-coupling terms, with nonlinear self-interactions
or density-dependent coupling strengths used to encode its medium dependence.
When combined with the Bogoliubov treatment of pairing and multidimensional
constraints, CDFT provides a natural basis for exploring reflection-asymmetric
configurations and extended fission pathways
\cite{Lu2012,Lu2014,Zhou2016,Zhao2017}. In the present work, we employ its
nonlinear point-coupling realization within the MDC-RHB model.

The Lagrangian density of the relativistic mean-field (RMF) model
with nonlinear point coupling (NL-PC) \cite{Lu2014,Zhou2016}
can be written as
\begin{equation}
  \mathcal{L}=\bar\psi(i\gamma_\mu\partial^\mu-M)\psi
  -\mathcal{L}_{\mathrm{4f}}-\mathcal{L}_{\mathrm{nl}}
  -\mathcal{L}_{\mathrm{der}}-\mathcal{L}_{\mathrm{em}},
  \label{eq:lagrangian}
\end{equation}
where
\begin{equation}
\begin{aligned}[b]
 \mathcal{L}_{\mathrm{4f}}
 &=\frac12\alpha_S\rho_S^2+\frac12\alpha_V\rho_V^2
   +\frac12\alpha_{TS}\vec\rho_{TS}^{\,2}
   +\frac12\alpha_{TV}\vec\rho_{TV}^{\,2},\\
 \mathcal{L}_{\mathrm{nl}}
 &=\frac13\beta_S\rho_S^3+\frac14\gamma_S\rho_S^4
   +\frac14\gamma_V[\rho_V^2]^2,\\
 \mathcal{L}_{\mathrm{der}}
 &=\frac12\delta_S[\partial_\nu\rho_S]^2
   +\frac12\delta_V[\partial_\nu\rho_V]^2
   +\frac12\delta_{TS}[\partial_\nu\vec\rho_{TS}]^2\\
 &\quad+\frac12\delta_{TV}[\partial_\nu\vec\rho_{TV}]^2,\\
 \mathcal{L}_{\mathrm{em}}
 &=\frac14 F^{\mu\nu}F_{\mu\nu}
   +e\frac{1-\tau_3}{2}A_0\rho_V
\end{aligned}
\label{eq:lagrangian_terms}
\end{equation}
are the four-fermion, nonlinear, derivative, and electromagnetic
terms, respectively. Here \(\psi\) is the nucleon field, \(M\) is
the nucleon mass, and
\(F_{\mu\nu}=\partial_\mu A_\nu-\partial_\nu A_\mu\).
The coefficients \(\alpha\), \(\beta\), \(\gamma\), and \(\delta\)
are the corresponding coupling constants. Arrows denote vectors in
isospin space, with \(\vec\tau\) the Pauli matrices.
The above notation applies to static, time-reversal-invariant states,
for which only the time-like components of the vector currents survive:
\begin{equation}
\begin{aligned}
 \rho_S&=\bar\psi\psi,&
 \rho_V&=\bar\psi\gamma^0\psi,\\
 \vec\rho_{TS}&=\bar\psi\vec\tau\psi,&
 \vec\rho_{TV}&=\bar\psi\gamma^0\vec\tau\psi .
\end{aligned}
\label{eq:bilinears}
\end{equation}
In the Hartree and no-sea approximations, the bilinears are replaced by
their expectation values in the quasiparticle vacuum.
Without proton-neutron mixing, only the third isospin component of
the isovector densities survives, denoted by \(\rho_{TS}\) and
\(\rho_{TV}\). The charge projector \((1-\tau_3)/2\) in
\(\mathcal{L}_{\mathrm{em}}\) selects the proton density.
For the case \(\alpha_{TS}=\delta_{TS}=0\) used here
(see Sec.~\ref{sec:numerical_details}), the scalar and vector potentials are
\begin{align}
 S(\bs r)
 &=\alpha_S\rho_S+\beta_S\rho_S^2+\gamma_S\rho_S^3
   +\delta_S\nabla^2\rho_S,
 \label{eq:scalar_potential}\\
 V(\bs r)
 &=\alpha_V\rho_V+\gamma_V\rho_V^3+\delta_V\nabla^2\rho_V
   \nonumber\\
 &\quad+\tau_3\bigl(\alpha_{TV}\rho_{TV}
             +\delta_{TV}\nabla^2\rho_{TV}\bigr)
   +e\frac{1-\tau_3}{2}A_0 .
 \label{eq:vector_potential}
\end{align}
All densities and fields in these expressions depend on \(\bs r\);
\(\tau_3=+1\) for neutrons and \(-1\) for protons.
The Coulomb field satisfies
\(-\nabla^2 A_0=e\rho_p\), with
\(\rho_p=(\rho_V-\rho_{TV})/2\).
The resulting single-particle Dirac Hamiltonian is
\begin{equation}
  \hat h_D=\bs\alpha\cdot\bs p+\beta[M+S(\bs r)]+V(\bs r).
  \label{eq:dirac}
\end{equation}
The derivation and implementation of the point-coupling fields are
given in Refs.~\cite{Lu2014,Zhou2016}.
\par\endgroup

\begingroup
\color{black}
\hypersetup{linkcolor=blue,citecolor=blue}
Pairing correlations are incorporated through the Bogoliubov
transformation \cite{Zhao2017}. In coordinate space, the RHB equation reads
\par\endgroup
\begin{equation}
  \int \mathrm{d}^3\bs r'\,
  \begin{pmatrix}
    h_D-\lambda & \Delta \\
    -\Delta^\ast & -h_D^\ast+\lambda
  \end{pmatrix}
  \binom{U_k}{V_k}
  =
  E_k\binom{U_k}{V_k},
  \label{eq:rhb}
\end{equation}
\begingroup
\color{black}
\hypersetup{linkcolor=blue,citecolor=blue}
where coordinate arguments have been suppressed.
Here \(\lambda\) is the chemical potential, \(\Delta\) is the pairing
field, \(E_k\) is the quasiparticle energy, and \((U_k,V_k)^T\) is the
quasiparticle wave function.
\par\endgroup

\begingroup
\color{black}
\hypersetup{linkcolor=blue,citecolor=blue}
At zero temperature, the local densities entering the potentials are
expressed in terms of the Bogoliubov amplitudes \(V_k\) as
\cite{Zhao2017}
\begin{equation}
\begin{aligned}
 \rho_S(\bs r)&=\sum_{k>0}V_k^\dagger(\bs r)\gamma^0 V_k(\bs r),\\
 \rho_V(\bs r)&=\sum_{k>0}V_k^\dagger(\bs r)V_k(\bs r),\\
 \rho_{TV}(\bs r)&=\sum_{k>0}V_k^\dagger(\bs r)\tau_3 V_k(\bs r).
\end{aligned}
\label{eq:rhb_densities}
\end{equation}
Here \(V_k\) is the hole component of the Bogoliubov wave function,
including both large and small Dirac components. The sums run over the
positive-energy nucleonic quasiparticle states retained in the no-sea
approximation, including both members of each time-reversed pair;
therefore, no additional factor of two is required.
\par\endgroup

\begingroup
\color{black}
\hypersetup{linkcolor=blue,citecolor=blue}
The pairing field is generated from the finite-range separable force
\cite{Tian2009,Tian2009RQRPA,Tian2009ADHO},
\par\endgroup
\begin{equation}
  {\color{black}V}
  =
  -G\,\delta(\bs {\rm R}-\bs {\rm R}')
  P(\bs r)P(\bs r')
  \frac{1-P^\sigma}{2},
  \label{eq:pairing}
\end{equation}
with
\begin{equation}
  P(\bs r)=\frac{1}{(4\pi a^2)^{3/2}}
  \exp\left(-\frac{\bs r^2}{4a^2}\right).
  \label{eq:pairingform}
\end{equation}
\begingroup
\color{black}
\hypersetup{linkcolor=blue,citecolor=blue}
Here \(\bs {\rm R}=(\bs {\rm r}_1+\bs {\rm r}_2)/2\) and
\(\bs {\rm r}=\bs {\rm r}_1-\bs {\rm r}_2\) are the center-of-mass and
relative coordinates, \(P^\sigma\) exchanges the two spins, and the
primed coordinates refer to the final state of the pair.
The strength \(G\) and range \(a\) are specified in
Sec.~\ref{sec:numerical_details}.
\par\endgroup

\begingroup
\color{black}
\hypersetup{linkcolor=blue,citecolor=blue}
The deformation parameters associated with the constrained mass
multipole moments are
\begin{equation}
 \beta_{\lambda\mu}=\frac{4\pi}{3AR^\lambda}
 \langle\hat Q_{\lambda\mu}\rangle,\qquad
 \hat Q_{\lambda\mu}=r^\lambda Y_{\lambda\mu}(\theta,\phi),
 \label{eq:deformation}
\end{equation}
where \(R=1.2A^{1/3}\) fm and \(A\) is the mass number.
Within \(V_4\) symmetry, multipoles with even \(\mu\) can be included.
The modified linear constraint method minimizes
\begin{equation}
 E'=E_{\mathrm{RHB}}+\sum_{\lambda\mu}C_{\lambda\mu}\beta_{\lambda\mu},
 \label{eq:linearconstraint}
\end{equation}
with the multipliers \(C_{\lambda\mu}\) adjusted iteratively to reach the
target deformations \cite{Lu2014}. Algorithmic details and the basis
representation are given in Refs.~\cite{Lu2014,Zhao2017}.
For the collective coordinates \(q=(\beta_{20},\beta_{30})\) used below,
the potential energy is measured from the compact ground-state minimum:
\begin{equation}
 V_{\mathrm{PES}}(q)
 =E_{\mathrm{RHB}}(q)-E_{\mathrm{RHB}}(q_{\mathrm{g.s.}}).
 \label{eq:pes}
\end{equation}
\par\endgroup

\subsection{Finite-temperature BCS level density}
\label{sec:ftbcs}

\begingroup
\hypersetup{linkcolor=blue,citecolor=blue}
The master equation for shape evolution on the PES
requires the knowledge of intrinsic level densities and the excitation energy available.
At each constrained deformation \(q=(\beta_{20},\beta_{30})\), the
MDC-RHB solution supplies the single-particle energies
\(\varepsilon_k^\tau(q)\) and zero-temperature pairing gaps
\(\Delta_\tau(0,q)\), where \(\tau=\nu,\pi\) denotes neutrons or protons.
The intrinsic neutron and proton systems are treated with finite-temperature
BCS theory at a common, locally defined, deformation-dependent temperature \(T_q\)
\cite{Decowski1968,Adeev1989}. The temperature is fixed by the available
intrinsic excitation energy, as specified below. The single-particle
spectrum and self-consistent shape at each \(q\) are kept at their
zero-temperature MDC-RHB values; thermal readjustment of the underlying
mean field and PES is therefore neglected.
\par\endgroup

Defining the quasiparticle energies as
\begin{equation}
  E_k^\tau(T_q,q)
  =
  \left[
    \left(\varepsilon_k^\tau(q)-\lambda_\tau(T_q,q)\right)^2
    +\Delta_\tau^2(T_q,q)
  \right]^{1/2},
  \label{eq:qpenergy}
\end{equation}
the particle-number equation and the corresponding gap equation read
\begingroup
\color{black}
\hypersetup{linkcolor=blue,citecolor=blue}
\begin{align}
 N_\tau
 &=\sum_{k>0}\left[
  1-\frac{\varepsilon_k^\tau(q)-\lambda_\tau(T_q,q)}
  {E_k^\tau(T_q,q)}
  \tanh\left(\frac{E_k^\tau(T_q,q)}{2T_q}\right)
  \right],
 \label{eq:number}\\
 \frac{2}{G_\tau(q)}
 &=\sum_{k>0}\frac{1}{E_k^\tau(T_q,q)}
  \tanh\left(\frac{E_k^\tau(T_q,q)}{2T_q}\right).
 \label{eq:gap}
\end{align}
Here \(N_\nu=N\), \(N_\pi=Z\), and \(k>0\) selects one state from
each time-reversed pair. The effective strengths \(G_\tau(q)\) are
adjusted to reproduce the zero-temperature MDC-RHB gaps as described in
Sec.~\ref{sec:numerical_details}. Equations~(\ref{eq:number}) and
(\ref{eq:gap}) determine the chemical potentials \(\lambda_\tau\) and
pairing gaps \(\Delta_\tau\). Above the BCS critical temperature,
\(\Delta_\tau=0\), and only the particle-number equation is solved.
The quantity in square brackets in Eq.~(\ref{eq:number}), denoted by
\(n_k^\tau(T_q,q)\), represents the total occupation of a time-reversed pair, \(\left(0\leq n_k^\tau\leq2\right)\).
\par\endgroup

The internal energy of each neutron or proton subsystem is
\begin{equation}
  {\cal E}_\tau(T_q,q)
  =
  \sum_{k>0}\varepsilon_k^\tau(q)n_k^\tau(T_q,q)
  -\frac{\Delta_\tau^2(T_q,q)}{G_\tau(q)}
  \label{eq:thermalenergy}
\end{equation}
and its excitation energy is defined relative to the paired zero-temperature solution at the same deformation,
\begin{equation}
  U_\tau(T_q,q)
  =
  {\cal E}_\tau(T_q,q)-{\cal E}_\tau(0,q).
  \label{eq:excitation}
\end{equation}
The entropy of each intrinsic subsystem is
\begin{equation}
\begin{aligned}
  S_\tau(T_q,q)
  &=
  2\sum_{k>0}
  \left[
    \ln\left(1+\mathrm{e}^{-E_k^\tau(T_q,q)/T_q}\right)
  \right.
  \\
  &\quad
  \left.
    +\frac{E_k^\tau(T_q,q)/T_q}
    {1+\mathrm{e}^{E_k^\tau(T_q,q)/T_q}}
  \right].
\end{aligned}
  \label{eq:entropy}
\end{equation}
Neglecting explicit neutron-proton pairing, the total intrinsic excitation energy and entropy are additive,
\begin{eqnarray}
    U(T_q,q)=U_\nu(T_q,q)+U_\pi(T_q,q),
  \nonumber \\
  S(T_q,q)=S_\nu(T_q,q)+S_\pi(T_q,q).
  \label{eq:additiveentropy}  
\end{eqnarray}

At each deformation $q$, the local temperature $T_q$ is determined by
requiring the intrinsic excitation energy $U(T_q,q)$ to satisfy energy
conservation. With the compact ground-state minimum chosen as the energy
reference, the total energy of the fissioning nucleus is
$E_{\mathrm{g.s.}}^{\mathrm{comp}}+E^*$, where $E^*$ denotes the excitation
energy above this minimum.
The energy-conservation condition is
\begin{equation}
  U(T_q,q)
  =
  E^\ast-\left[E(q)-E_{\mathrm{g.s.}}^{\mathrm{comp}}\right].
  \label{eq:localtemperature}
\end{equation}
Here $E(q)$ is the zero-temperature MDC-RHB energy of the constrained
configuration. The right-hand side gives the intrinsic excitation energy
available at $q$ and determines the local temperature $T_q$. Configurations
for which this quantity is negative are energetically inaccessible.

It is important to note that \(E^\ast\) is the conserved total excitation energy and does not correspond to a uniform temperature over the whole collective space. At each deformation, part of the total energy is stored in the potential-energy increase $E(q)-E_{\mathrm{g.s.}}^{\mathrm{comp}}$, and only the remainder is available for intrinsic excitation. Consequently, the local temperature $T_q$ varies along the PES.

The intrinsic level density at fixed excitation energy, neutron number,
proton number, and deformation is obtained from the inverse Laplace
transform of the grand-canonical partition function in the saddle-point
approximation \cite{Adeev1989,Decowski1968},
\begin{equation}
  \rho_{\mathrm{int}}(U,N,Z;q)
  =
  \frac{\exp[S(U,N,Z;q)]}{(2\pi)^{3/2}{\color{black}\cal D}},
  \label{eq:leveldensity}
\end{equation}
where \({\cal D}\) accounts for Gaussian fluctuations around the saddle point.
Introducing \(\beta_q=1/T_q\) and
\(\alpha_\tau=\beta_q\lambda_\tau(T_q,q)\), this factor is evaluated from the
thermodynamic curvature matrix
\begin{equation}
  {\cal D}
  =
  \left[
  \det
  \begin{pmatrix}
  S_{\beta_q\beta_q} & S_{\beta_q\alpha_\nu} & S_{\beta_q\alpha_\pi}\\
  S_{\alpha_\nu\beta_q} & S_{\alpha_\nu\alpha_\nu} & 0\\
  S_{\alpha_\pi\beta_q} & 0 & S_{\alpha_\pi\alpha_\pi}
  \end{pmatrix}
  \right]_{\mathrm{sp}}^{1/2},
  \label{eq:fluctuation}
\end{equation}
with \(S_{xy}\equiv\partial^2S/(\partial x\,\partial y)\).
The mixed neutron-proton derivative vanishes because the neutron and proton quasiparticle systems are independent
apart from their common temperature.
Finally, the local temperature is eliminated in favor of \(U\), giving
\(S(U,N,Z;q)\) and \(\rho_{\mathrm{int}}(U,N,Z;q)\).

For the numerical implementation, the collective space is discretized
into lattice points $i\equiv q_i$, with
$q_i=(\beta_{20}^{(i)},\beta_{30}^{(i)})$. At each mesh point and  intrinsic excitation
energy  $ U_i \equiv U(T_{q_i},q_i)$ the corresponding
level density is denoted by
$
  \rho_i
  \equiv
  \rho_{\mathrm{int}}(U_i,N,Z;q_i).
  \label{eq:meshquantities}
$
For $U_i<0$, the configuration is energetically inaccessible and
$\rho_i=0$.
This construction follows the finite-temperature microscopic level-density formalism used in recent applications to
heavy and superheavy nuclei \cite{Rahmatinejad2021,Rahmatinejad2022,Rahmatinejad2024},
but here the deformation-dependent microscopic input is taken from the MDC-RHB solutions described above.
The deformation dependence of shell gaps, level bunching, and pairing correlations enters directly through the
MDC-RHB single-particle spectra.
The quantity $\rho_i$ is an intrinsic state density;
spin-parity projection and rotational or vibrational collective enhancement factors are not included in the present calculation.

\subsection{Master-equation dynamics and scission probabilities}
\label{sec:master_equation}
The shape evolution toward scission is described on the discretized
collective mesh by a Markovian master equation. Such a statistical
transport description is appropriate for strongly damped collective
motion, for which the evolution can be regarded as stochastic transitions
between neighboring nuclear shapes
\cite{RandrupMoller2011,RandrupMollerSierk2011,Ward2017}.
In contrast to Langevin approaches, which require collective inertias and
friction coefficients, the present formulation propagates directly the
occupation probabilities on the microscopic deformation mesh and allows
the deformation-dependent level densities to enter explicitly into the
transition rates.

The probability
$P(i,t)$ to occupy the configuration $q_i$ evolves according to
\begin{equation}
  \frac{\mathrm{d}P(i,t)}{\mathrm{d}t}
  =
  \sum_j
  \left[
    \Lambda_{ij}P(j,t)-\Lambda_{ji}P(i,t)
  \right],
  \label{eq:master}
\end{equation}
where $\Lambda_{ij}$ denotes the transition
rate from $q_j$ to $q_i$.
For neighboring points on the deformation mesh, the transition rate is
written in the Fermi golden-rule form:
\begin{equation}
  \Lambda_{ij}
  =
  \frac{2\pi}{\hbar}
  \left|\langle i|\hat V|j\rangle\right|^2
  \rho_i
  \equiv
  \lambda_{ij}\rho_i ,
  \label{eq:transitionrate}
\end{equation}
where $\rho_i$ is the intrinsic state density at the final configuration
$q_i$ introduced above. Thus, the deformation-dependent microscopic
level density determines the available final-state phase space for the
shape transition.

Following the microscopic transition-probability prescription of Refs.~\cite{MorettoSventek1975,Adamian1994},
we use
\begin{align}
  \lambda_{ij}
  &=
  \frac{\lambda}{\sqrt{\rho_i\rho_j}},
  \nonumber\\
  \Lambda_{ij}
  &=
  \lambda
  \sqrt{\frac{\rho_i}{\rho_j}},
  \label{eq:microscopic_rate}
\end{align}
for two connected energetically available shapes. This prescription automatically satisfies the detailed-balance condition
$\Lambda_{ij}\rho_j=\Lambda_{ji}\rho_i$.
\begingroup
\color{black}
\hypersetup{linkcolor=blue,citecolor=blue}
Here, $\lambda$ sets the common microscopic transition-rate scale and is
taken to be the same for changes in the quadrupole and octupole deformations.
\par\endgroup

The evolution described by Eq.~(\ref{eq:master}) is continued until the
system reaches a mesh point belonging to the scission line
$\mathcal{S}$. This line defines the boundary in the collective space
at which the fissioning nucleus is assumed to separate into
two fragments. In the master-equation treatment, $\mathcal{S}$ is
therefore taken as an absorbing boundary: once the system reaches this
boundary, it is assumed to undergo fission. The probability
associated with a scission point $i_s\in\mathcal{S}$ is obtained by integrating
the probability current entering that point,
\begin{equation}
  \mathcal{P}(i_s)
  =
  \int_{0}^{\infty}
  \sum_{j\notin\mathcal{S}}
  \Lambda_{i_sj}P(j,t)\,\mathrm{d}t .
  \label{eq:fissionyield}
\end{equation}

To elucidate the influence of the scission-line properties relative
to that of the preceding shape evolution, we also consider a separate
equilibrium prescription in which the relative populations of
scission configurations are determined by their intrinsic state
densities.
The corresponding probability distribution is then determined directly
by the statistical weights of scission configurations:
\begin{align}
  P^{\mathrm{eq}}(i_s)
  &=
  \frac{\rho_{i_s}}
       {\displaystyle\sum_{i'_s\in\mathcal{S}}\rho_{i'_s}},
  \nonumber\\
  \frac{P^{\mathrm{eq}}(i_s)}
       {P^{\mathrm{eq}}(i'_s)}
  &=
  \frac{\rho_{i_s}}{\rho_{i'_s}} .
  \label{eq:equilibriumweight}
\end{align}
\begingroup
\color{black}
\hypersetup{linkcolor=blue,citecolor=blue}
The numerical sampling is specified in Sec.~\ref{sec:numerical_details}.
The geometrical scission criterion and the extraction of prefragment
properties from the densities are described in Sec.~\ref{sec:pes_scission}.
\par\endgroup

\section{Results and Discussion}
\label{sec:results}

\subsection{Numerical details}
\label{sec:numerical_details}

\begingroup
\color{black}
\hypersetup{linkcolor=blue,citecolor=blue}
The particle-hole channel is described by the nonlinear point-coupling
functional PC-PK1 \cite{Zhao2010}, for which the isovector-scalar
couplings vanish, \(\alpha_{TS}=\delta_{TS}=0\).
In the RHB pairing channel, the
finite-range separable force has \(G=728.0\) MeV fm\(^3\) and
\(a=0.644\) fm. These parameters reproduce the
\({}^{1}S_0\) pairing gap in nuclear matter calculated with the Gogny
D1S interaction \cite{Tian2009,Tian2009ADHO}.
The calculations impose axial and time-reversal symmetries while
allowing reflection symmetry to break. The constrained collective
coordinates are \((\beta_{20},\beta_{30})\); a dipole constraint
\(\beta_{10}=0\) fixes the center of mass, and the remaining allowed
multipoles are determined variationally.
\par\endgroup

The RHB equations are solved by expanding the quasiparticle wave functions in an axially deformed
harmonic-oscillator (ADHO) basis \cite{Gambhir1990_APNY198-132}.
For the Dirac spinors, all fermionic major shells up to \(N_{\mathrm{F}}=20\) are included.
The densities and mean fields are evaluated on Gauss-Hermite and Gauss-Laguerre spatial meshes with \(N_z=36\) and \(N_\rho=18\).
Both positive and negative \(z\) mesh points are retained, which is essential for reflection-asymmetric shapes.

The zero-temperature PES of \Hg{} is calculated in the two-dimensional collective space \((\beta_{20},\beta_{30})\).
The constrained grid covers \(-0.52\leq\beta_{20}\leq4.88\) with a mesh spacing of 0.04 and
\(0.00\leq\beta_{30}\leq3.00\) with a mesh spacing of 0.05.
At each point the RHB equations, densities, pairing tensor, mean fields,
and constraint terms are iterated self-consistently with Broyden mixing.
The deformation energy is normalized to the compact ground-state minimum according to Eq.~(\ref{eq:pes}).
The resulting PES describes the potential-energy evolution with \(\beta_{20}\) and \(\beta_{30}\).

\begingroup
\color{black}
\hypersetup{linkcolor=blue,citecolor=blue}
For the subsequent finite-temperature BCS calculation, the effective
strengths \(G_\nu(q)\) and \(G_\pi(q)\) are determined separately at each
deformation to reproduce the neutron and proton pairing gaps of the
zero-temperature MDC-RHB solution.
The pairing space includes 200 pairs of time-reversed single-particle states for each nucleon species, and this number is kept fixed at all deformations.

The intrinsic state densities obtained from the finite-temperature
BCS calculation are then used to evaluate the transition rates
entering the master equation according to
Eq.~(\ref{eq:microscopic_rate}). 
Stochastic transitions connect nearest-neighbor mesh points
along the $\beta_{20}$ and $\beta_{30}$ directions.
Since the overall transition-rate scale $\lambda$ determines only the time
scale of the evolution without affecting the resulting scission
probabilities or fragment yields, we set $\lambda=1$ in the following.

\par\endgroup
A set of 100 initial configurations is sampled from the compact
ground-state region of the deformation mesh. The candidate configurations
are taken from the connected low-energy basin surrounding the ground-state
minimum and are selected with probabilities proportional to the
corresponding microscopic level densities $\rho_i$. From each selected
initial configuration, 1000 stochastic trajectories are propagated until
the scission boundary is reached. Increasing either the number of sampled
initial configurations or the number of trajectories per configuration
does not produce any noticeable change in the resulting scission
probabilities or fragment-mass distributions.

\subsection{Potential-energy surface and scission configurations}
\label{sec:pes_scission}

\begin{figure*}[tbp]
  \centering
  \includegraphics[width=0.86\textwidth]{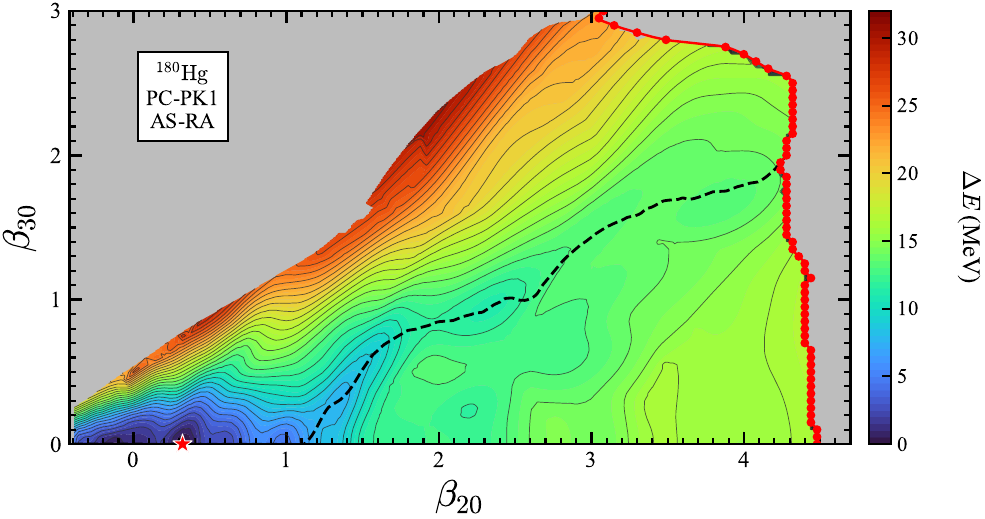}
  \caption{PES of \Hg{} in the \((\beta_{20},\beta_{30})\) plane calculated with the
  PC-PK1 density functional in the axially symmetric and reflection-asymmetric MDC-RHB model.
  The energies are given relative to the ground state according to Eq.~(\ref{eq:pes}).
  The black dashed line denotes the static fission path obtained by following the lowest connected valley,
  and the red circles connected by a solid line mark the candidate scission configurations selected using the
  neck criterion defined below. The red star marks the calculated ground-state minimum.}
  \label{fig:pes_scission}
\end{figure*}

\begingroup
\hypersetup{linkcolor=blue,citecolor=blue}
The PES in the \((\beta_{20},\beta_{30})\) plane (Fig.~\ref{fig:pes_scission}) exhibits a fission valley
extending from the compact region toward large quadrupole and finite octupole deformations.
At large elongations, reflection-symmetric configurations along \(\beta_{30}=0\) lie higher in energy and
form a ridge-like structure.
This preference for asymmetric evolution agrees with the SkM$^*$ and Gogny D1S HFB results of
Warda \textit{et al.} \cite{Warda2012}.
Ichikawa \textit{et al.} also found a ridge separating the saddle region from the symmetric
$^{90}$Zr+$^{90}$Zr fusion valley \cite{Ichikawa2012}, while the PC-PK1 calculations of
Li \textit{et al.} likewise favored asymmetric scission \cite{Li2022}.
The present topology therefore provides a static basis for asymmetric fission, consistent with the observed
\(A_{\mathrm{L}}/A_{\mathrm{H}}\simeq80/100\) mass split in \Hg{}
\cite{Andreyev2010,Elseviers2013}.
\par\endgroup

\begingroup
\color{black}
\hypersetup{linkcolor=blue,citecolor=blue}
\begin{samepage}
In the compact ground-state basin, the calculated lowest minimum corresponds to prolate shape with
\((\beta_{20},\beta_{30})=(0.32,0.00)\), while the lowest oblate solution near
\((\beta_{20},\beta_{30})=(-0.12,0.00)\) lies only about \(0.32\) MeV higher.
Experimentally, the ground-state band of \Hg{} is commonly associated with a weakly deformed oblate configuration,
coexisting with a more strongly deformed prolate structure
\cite{Dracoulis1988,Kondev2000,Heyde2011}.
Self-consistent mean-field calculations, however, frequently
place the minimum associated with the prolate configuration slightly below
that associated with the oblate configuration
\cite{Warda2010}.
The relative ordering of these shallow minima is sensitive to the energy-density functional, pairing correlations,
triaxial deformation, and collective correlations beyond the mean-field approximation.
The small calculated energy difference should therefore be regarded as evidence for competing low-lying configurations
rather than as a robust prediction of their ordering.
\par\end{samepage}

This ambiguity has little influence on the present fission calculation.
The fission-barrier energy is much larger than the energy difference between the two compact minima, and the subsequent
shape evolution enters the same outer region of the potential-energy surface.
We have verified that choosing either the prolate or the oblate configuration as the initial point leads to practically
identical scission probabilities and fragment-mass distributions.
\par\endgroup

\begin{figure}[!htbp]
  \centering
  \includegraphics[width=0.98\columnwidth]{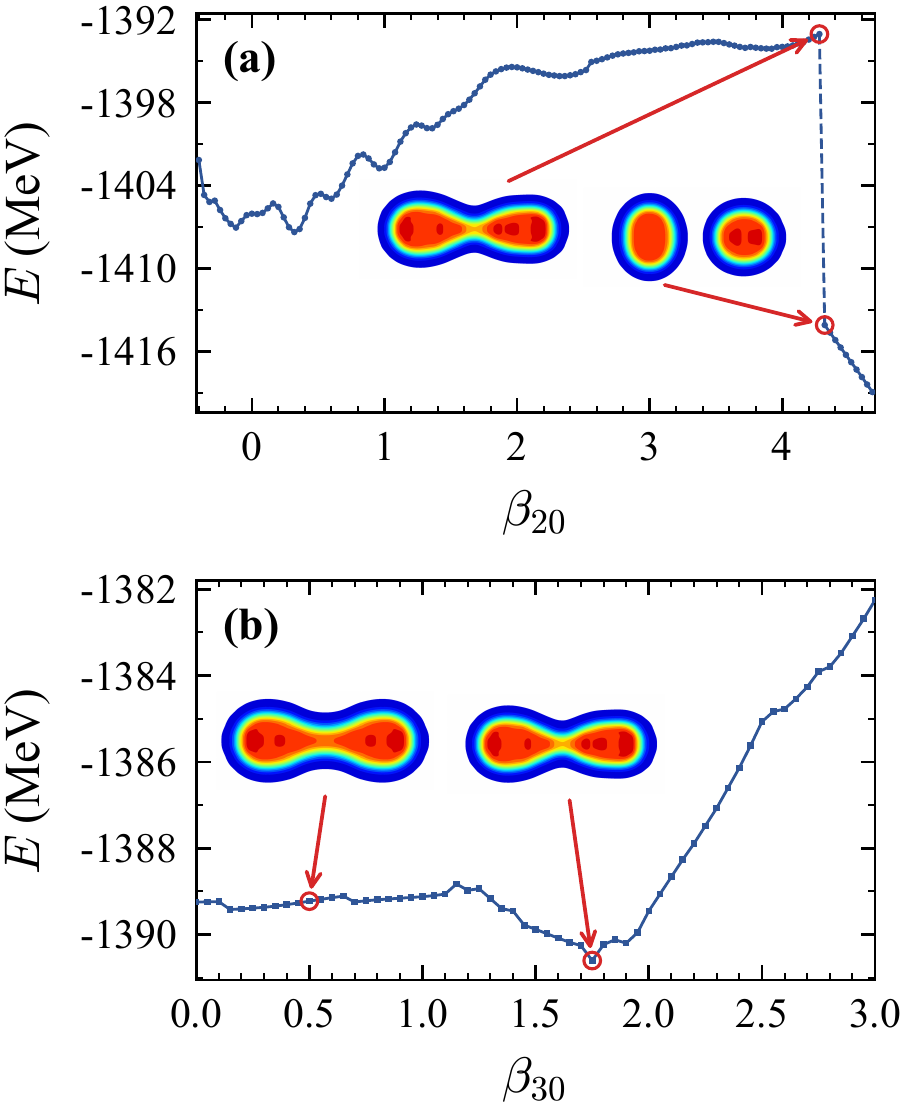}
  \caption{RHB total energies of \Hg{}:
  (a) along the static fission path versus \(\beta_{20}\) (blue circles)
  and (b) along the \(\rho=0.10~\mathrm{fm}^{-3}\) scission line versus
  \(\beta_{30}\) (blue squares). Red arrows indicate selected curve points
  at \(\beta_{20}=4.28\) and 4.32 in (a) and \(\beta_{30}=0.50\) and 1.75 in (b).
The insets show self-consistent density distributions that illustrate representative 
shapes and indicate their correspondence to the curve points.
  The dashed segment in (a) connects the two points bracketing the abrupt
  energy drop.}
  \label{fig:path_energy}
  \label{fig:scissionline_energy}
\end{figure}

To visualize the static fission path, we follow the lowest connected valley from the ground-state region toward
increasing elongation, as indicated by the black dashed line in Fig.~\ref{fig:pes_scission}.
At each value of \(\beta_{20}\), \(\beta_{30}\) is fixed at the value corresponding to the selected trajectory,
while the remaining unconstrained deformation degrees of freedom are determined by energy minimization.
The resulting one-dimensional potential is defined as:
\begin{equation}
  V_{\mathrm{path}}(\beta_{20})
  =
  V_{\mathrm{PES}}
  \left(\beta_{20},\beta_{30}^{\mathrm{path}}(\beta_{20})\right).
  \label{eq:path_energy}
\end{equation}
The corresponding RHB total energy is shown in the upper panel of Fig.~\ref{fig:path_energy}.
It represents an adiabatic reference path rather than a real-time trajectory.

The path energy \(V_{\mathrm{path}}\) increases along the sequence of
one-body configurations that retain a finite neck and reaches its highest
value in the region where the neck starts to decrease rapidly. Beyond this
region, the static energy drops sharply as the density develops a pronounced
two-fragment character. Thus, within the resolution of the calculated
deformation mesh, no distinct barrier maximum followed by an extended
downhill evolution toward scission is obtained. The effective barrier top
within the present collective space is therefore located very close to the
scission region.

This behavior is not restricted to the selected one-dimensional path.
Within the calculated \((\beta_{20},\beta_{30})\) space, the energetically
accessible paths generally rise toward the same region, indicating a close
proximity between the barrier top and scission. Such a short
barrier-to-scission interval is characteristic of pre-actinide nuclei,
whose transition-state configurations are generally strongly elongated
and close to scission, in contrast to the longer saddle-to-scission
descent typical of actinides and superheavy nuclei
\cite{Bogachev2021,Bernard2024}.

\begin{figure}[!htbp]
  \centering
  \includegraphics[width=0.78\columnwidth]{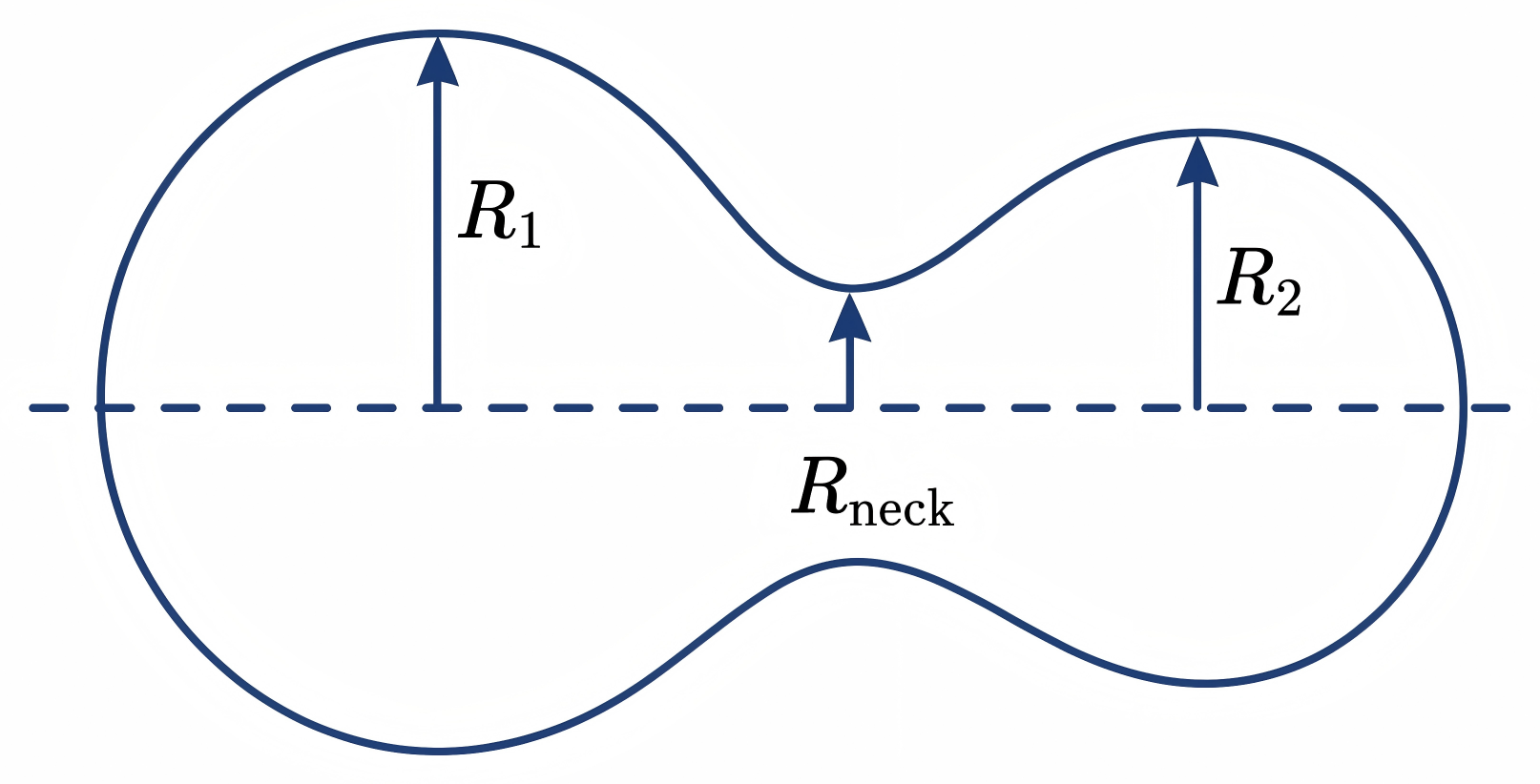}
  \caption{Schematic definition of the geometric shape of a reflection-asymmetric scission point.
  The radii \(R_1\) and \(R_2\) characterize the two nascent fragments on the density contour, while
  \(R_{\mathrm{neck}}\) denotes the minimum radial extension in the neck region.}
  \label{fig:scission_configuration}
\end{figure}

The rapid reduction of the neck in the pre-actinide mass region provides a natural
basis for defining the scission boundary geometrically. Near the abrupt decrease of the static path energy shown
in Fig.~\ref{fig:path_energy}(a), the neck becomes strongly reduced.
We identify scission configurations using a geometrical criterion applied
to the self-consistent density distribution. The contour
$\rho(r,z)=0.10~\mathrm{fm}^{-3}$ defines the nuclear surface and the neck
radius $R_{\mathrm{neck}}$ is the minimum radial extension of this surface
in the neck region between the two nascent fragments
(Fig.~\ref{fig:scission_configuration}). A configuration is identified
as a scission configuration when
\begin{equation}
  R_{\mathrm{neck}}\leq R_{\mathrm{crit}},
  \qquad
  R_{\mathrm{crit}}=1.0~\mathrm{fm}.
  \label{eq:scissioncriterion}
\end{equation}
This value provides a practical criterion for the transition from a
strongly necked one-body configuration to a nearly separated binary
system and is consistent with neck-radius criteria commonly employed
in fission calculations \cite{Kostryukov2021,Ishizuka2023,Han2021}. Larger values of $R_{\rm crit}$ would terminate the evolution while
the system still retains a pronounced one-body character, whereas at
smaller neck radii the potential energy decreases rapidly and the
present collective description becomes less suitable.

This geometrical definition is particularly appropriate for
$^{180}$Hg, where the emergence of a well-developed binary configuration
is confined to a relatively narrow region near scission. This differs
from heavier fissioning systems, where fragment-like structures may
develop over a broader range of elongations and continue to evolve within
an extended pre-scission binary region. In such systems, the geometrically
defined scission line represents mainly the outer boundary of this region
rather than the onset of binary fragment formation. In $^{180}$Hg, by
contrast, it provides a comparatively direct characterization of the
nascent-fragment properties.

For each fixed value of $\beta_{30}$, the scission point is taken as the
first configuration encountered with increasing $\beta_{20}$ that satisfies
Eq.~(\ref{eq:scissioncriterion}). The set of configurations selected in
this way defines the scission line in the
$(\beta_{20},\beta_{30})$ plane. Within the resolution of the deformation
mesh, this line lies in the same region where the static path energy drops
rapidly and the neck becomes strongly reduced, providing a consistency
check for the adopted geometrical definition. The geometrical definition of the scission configuration and the neck radius \(R_{\mathrm{neck}}\) is illustrated schematically in
Fig.~\ref{fig:scission_configuration}.

The neck criterion used here differs from approaches in
which a neck operator or neck coordinate $Q_{\mathrm{N}}$ is introduced
as an additional constrained collective coordinate and used directly to
identify scission configurations \cite{Li2022}. In the present approach,
$R_{\mathrm{neck}}$ provides a geometrical, coordinate-space criterion
for identifying the scission boundary and does not prescribe how the
particles in the neck region are assigned to the two nascent fragments.
This distinction is particularly important for diffuse neck configurations,
for which the particle partition is not uniquely fixed by the geometrical
scission criterion. The particle-number partition is therefore treated
as a separate step, as described below.

The lower panel of Fig.~\ref{fig:scissionline_energy} shows the energy along the scission line as a
function of $\beta_{30}$. A pronounced minimum occurs at finite octupole
deformation $\beta_{30}$ = 1.75, while the energy
increases toward both the nearly reflection-symmetric and the more
strongly asymmetric configurations. Thus, the scission line itself
exhibits an energetic preference for reflection-asymmetric configurations.

The minimum of the scission-line energy reflects the combined microscopic structure and
self-consistent deformation of the nascent fragments. Their shell and
pairing properties, Coulomb interaction, and deformation energy all
contribute to the scission energy. In particular, the reduction of the
Coulomb repulsion with increasing charge asymmetry can favor asymmetric
configurations, while the self-consistent deformation of the nascent
fragments incorporates their different energetic responses to shape
changes. The importance of fragment stiffness has also been emphasized
in improved scission-point calculations of Hg isotopes, where an explicit
zero-point-vibration contribution was included and found to enhance the
preference for asymmetric splitting around $^{180}$Hg
\cite{Andreev2013}. In the present MDC-RHB calculation, no separate
zero-point-vibration term is introduced; instead, the deformation
energy and its variation with shape are obtained self-consistently.
The interplay of these effects provides a microscopic basis for the
minimum of the scission-line energy at finite $\beta_{30}$.

For each configuration on the scission line, the fragment masses and
charges are extracted from the self-consistent neutron, proton, and total
nucleon densities $\varrho_\alpha(r,z)$
$(\alpha=n, p,A)$. Under axial symmetry, these
densities are integrated over the radial coordinate to obtain the
corresponding one-dimensional densities $\varrho_\alpha(z)$ along the
fission axis.

The particle-number partition of each scission configuration is determined
from a dividing point marked with $z_{\rm div}$, obtained from a Gaussian-filtered profile of
the total line density. Physically, this procedure locates the region of
minimum particle content between the two nascent fragments while averaging
over the finite spatial extent of the neck. Following the density-based
prescription of Ref.~\cite{YounesGogny2009}, we define
\begin{align}
  \mathcal{N}(z)
  &=
  \int_{z_{\min}}^{z_{\max}}
  \exp\left[-\frac{(z'-z)^2}{a_{\mathrm{neck}}^2}\right]
  \varrho_{\mathrm{A}}(z')\,\mathrm{d}z',
  \label{eq:neck_function}
\end{align}
with the Gaussian width $a_{\rm neck}=1.0~\mathrm{fm}$ as adopted in that work.
The function \(\mathcal{N}(z)\) represents
the density integrated over a finite region centered at \(z\), with a
characteristic longitudinal scale of approximately
\(2a_{\mathrm{neck}}\). 

The dividing coordinate $z_{\rm div}$ is
determined by requiring $\mathcal{N}(z)$ to attain a local minimum between the two fragment-density maxima:
\begin{align}
  \left.
  \frac{\mathrm{d}\mathcal{N}(z)}{\mathrm{d}z}
  \right|_{z=z_{\mathrm{div}}}
  &=0,
  \qquad
  \left.
  \frac{\mathrm{d}^2\mathcal{N}(z)}{\mathrm{d}z^2}
  \right|_{z=z_{\mathrm{div}}}>0 .
  \label{eq:zdiv_condition}
\end{align}
The neutron and proton numbers of the left (\(F=\mathrm L\)) and right
(\(F=\mathrm R\)) prefragments are obtained by integrating the
corresponding one-dimensional density distributions on either side of \(z_{\mathrm{div}}\),
\begin{equation}
  N_F=\int_{\Omega_F}\varrho_{\mathrm n}(z)\,\mathrm{d}z,
  \qquad
  Z_F=\int_{\Omega_F}\varrho_{\mathrm p}(z)\,\mathrm{d}z,
  \label{eq:fragment_NZA}
\end{equation}
where
\(\Omega_{\mathrm L}=[z_{\min},z_{\mathrm{div}}]\) and
\(\Omega_{\mathrm R}=[z_{\mathrm{div}},z_{\max}]\), with
\(A_F=N_F+Z_F\).

To calculate the intrinsic multipole deformations, each prefragment is
treated in its own center-of-mass frame. 
The origins of the corresponding
frames are placed at the axial centers of mass of the prefragments:
\begin{equation}
  z_F
  =
  \frac{1}{A_F}
  \int_{\Omega_F^{(3)}}
  z\,\varrho_{\mathrm A}(\mathbf r)\,
  \mathrm{d}^3\mathbf r ,
  \qquad (F=\mathrm L,\mathrm R).
  \label{eq:fragment_center}
\end{equation}
Let $\mathbf r$ denote a point in the global coordinate system
and let $\Omega_F^{(3)}$ be the three-dimensional region on the corresponding
side of $z_{\mathrm{div}}$. Define
$\mathbf r_F=\mathcal R_F(\mathbf r-z_F\hat{\mathbf z})$, where
$\mathcal R_F$ orients the local longitudinal axis outward from the neck.
The spherical coordinates of $\mathbf r_F$ are
$(r_F,\theta_F,\phi_F)$. The intrinsic mass multipole moments are
\begin{equation}
  Q_{\lambda0}^{F}
  =\int_{\Omega_F^{(3)}}
   \varrho_{\mathrm A}(\mathbf r)\,
   r_F^\lambda Y_{\lambda0}(\theta_F)\,
   \mathrm{d}^3\mathbf r ,
   \quad (\lambda=2,3).
  \label{eq:fragment_multipoles}
\end{equation}
The corresponding dimensionless quadrupole $\beta_{20}^{F}$ and octupole $\beta_{30}^{F}$ prefragment deformations are
defined by Eq.~\eqref{eq:deformation}, with $A$ and $R$ replaced by
$A_F$ and $R_F=1.2A_F^{1/3}$ fm, respectively.

Because the coordinate-space densities are integrated over a finite
numerical domain, the resulting neutron and proton numbers can differ
slightly from their exact values due to the omitted density tails and
numerical integration. We therefore apply spatially uniform normalization
factors such that \(A_{\mathrm L}+A_{\mathrm R}=180\) and
\(Z_{\mathrm L}+Z_{\mathrm R}=80\). The same mass-density normalization
is applied to \(Q_{\lambda0}^{F}\).
For the octupole deformation, the local axis is chosen to point outward
from the neck for both prefragments, avoiding an artificial sign difference
associated with their opposite locations. Accordingly, the octupole deformation is characterized by its magnitude $|\beta_{30}^{F}|$.

\begin{figure}[!htbp]
  \centering
  \includegraphics[width=\columnwidth]
{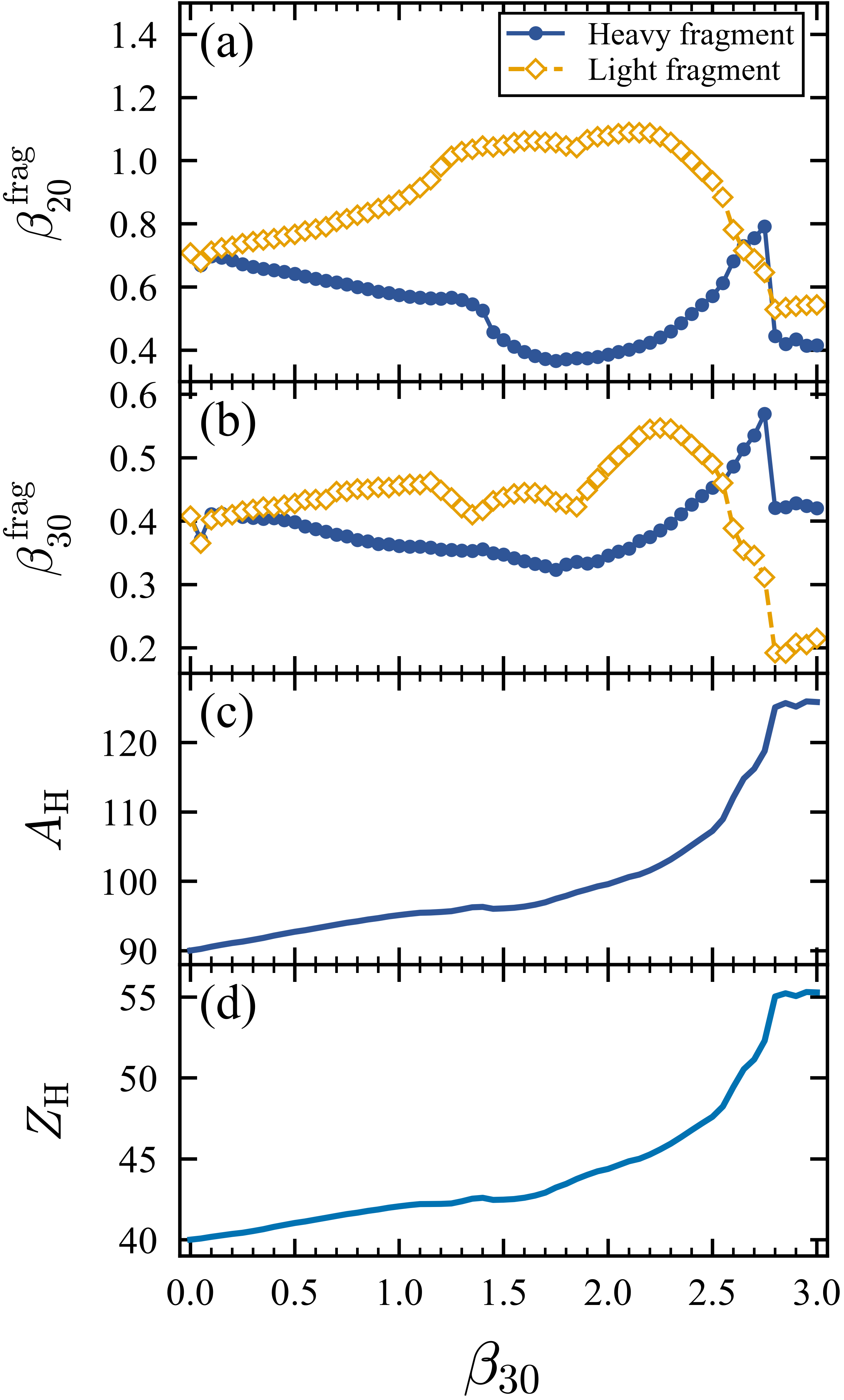}
\caption{Prefragment properties along the \(\rho=0.10~\mathrm{fm}^{-3}\) scission line:
  (a) and (b) show quadrupole \(\beta_{20}^{\mathrm{frag}}\) and octu\-po\-le \(\beta_{30}^{\mathrm{frag}}\) 
  deformations of pre-scission fragments, respectively. Blue filled circles with solid lines denote heavy prefragments; orange open diamonds with
  dashed lines denote light prefragments.
  Masses \(A_{\mathrm{H}}\) and charges \(Z_{\mathrm{H}}\) of heavy fragments are shown in (c) and (d), respectively.}
  \label{fig:fragment_properties}
\end{figure}

Along the selected scission line, each value of \(\beta_{30}\) corresponds
to one self-consistent density and therefore to one
\(z_{\mathrm{div}}\)-based partition into two prefragments. The resulting
prefragment properties are summarized in
Fig.~\ref{fig:fragment_properties}. 
The heavy-prefragment mass and charge
evolve from \(A_{\mathrm H}\simeq90\) and \(Z_{\mathrm H}\simeq40\) in
the nearly symmetric region toward larger asymmetry. Around the minimum
of the scission-line energy, \(\beta_{30}\simeq1.7\)--1.8, the heavy
prefragment has \(A_{\mathrm H}\approx97\)--98 and
\(Z_{\mathrm H}\approx43\). Toward the extreme-asymmetric end of the
scission line, \(A_{\mathrm H}\) and \(Z_{\mathrm H}\) increase rapidly
and then approach saturation. The accompanying abrupt changes of the
extracted prefragment deformations reflect the discrete selection of
scission configurations on the deformation mesh and occur in the same
region where the fragment partition changes rapidly. Since these
configurations lie considerably higher in energy, they are not expected
to contribute appreciably to the dominant fission yield.
In the energetically relevant asymmetric region, both prefragments remain
strongly deformed at scission, with the light prefragment generally more
elongated than the heavy one. Their sizable quadrupole and octupole
deformations show that the relevant scission configurations cannot be
represented by two cold spherical fragments. The observed
\(80/100\) mass split should therefore be associated with the structure
of strongly deformed nascent fragments and the energetically favored 
asymmetric scission configurations, rather than with the spherical
\(\Zr+\Zr\) limit.

\subsection{Fragment mass distributions}
The fragment-mass distribution is obtained by assigning the probability
of each scission configuration to the corresponding light and heavy
prefragment masses. This probability is estimated from the fraction
of trajectories terminating at that point, corresponding to the
integrated probability flux in Eq.~(\ref{eq:fissionyield}).
The ensemble size and the number of trajectories are specified
in Sec.~\ref{sec:numerical_details}.

\begin{figure}[!htbp]
  \centering
  \includegraphics[width=\columnwidth]{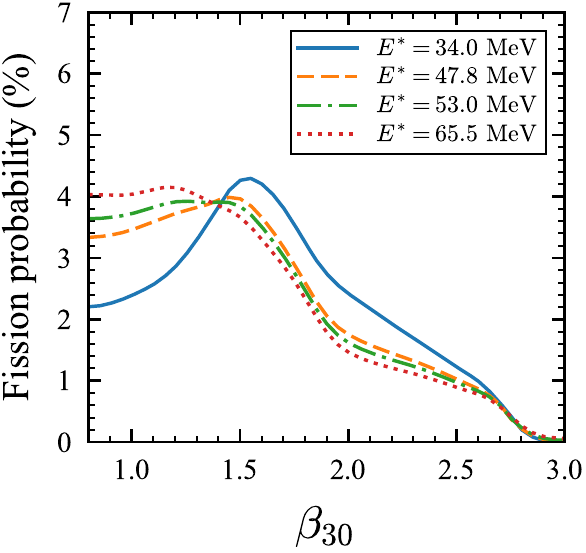}
  \caption{Calculated fission probability along the reflection-asymmetric scission branch of \Hg{} as a function
  of the scission-line coordinate \(\beta_{30}\).
  The excitation energies \(E^\ast=34.00\), \(47.80\), \(53.00\), and \(65.50\) MeV are
  represented by blue solid, orange dashed, green dash-dotted, and red dotted curves, respectively.}
  \label{fig:scission_probability}
\end{figure}

Figure~\ref{fig:scission_probability}
shows the calculated scission probabilities at excitation energies of compound
nucleus $E^*=34.0$, 47.8, 53.0, and 65.5 MeV. These values are chosen to further
use for comparison with experimental mass distributions reported in Refs.~\cite{Kozulin2022,Nishio2015}.
For all excitation energies considered here, the probability remains concentrated on the reflection-asymmetric
part of the scission line. At the lowest energy, the maximum lies near \(\beta_{30}\simeq1.5\), close to the minimum of the scission-line
energy shown in Fig.~\ref{fig:scissionline_energy}. As the excitation energy increases, the distribution broadens and shifts moderately toward smaller octupole
deformation, but no dominant component develops near \(\beta_{30}=0\).
Thus the excitation-dependent level density modifies the relative weights of neighboring scission configurations
without removing the asymmetric preference imposed by the static landscape.

The enhancement of the symmetric component is most pronounced between
$E^*=34.0$ and $47.8$ MeV, whereas the changes at higher excitation
energies are weaker. At the lower excitation energy,
configurations along the nearly symmetric path have relatively little
intrinsic excitation available because of their higher potential energy
and are therefore strongly suppressed by the corresponding microscopic
level densities. As $E^*$ increases, these configurations acquire
substantially larger statistical weights. Once the symmetric path becomes
well accessible, a further increase of excitation energy affects the
competing paths more similarly, resulting in a weaker change of their
relative contributions.

Before relating the probability distribution in the
$(\beta_{20},\beta_{30})$ space to the fragment-mass distribution,
we examine how higher multipole shape degrees of freedom
that are not explicitly constrained in the present two-dimensional
calculation may affect
the width of the calculated mass distributions.
In the present multidimensionally constrained calculation, only the
quadrupole and octupole deformations are explicitly constrained when the
scission line is constructed, whereas the remaining multipole degrees of
freedom are determined variationally at their local minima.
Scanning an additional constrained multipole allows alternative
shapes and fragment pairs to be explored at fixed
$(\beta_{20},\beta_{30})$, beyond the particular configuration
selected by local energy minimization.
To estimate the resulting uncertainty, we perform an additional constrained
calculation in the hexadecapole direction for the representative scission
configuration
$(\beta_{20},\beta_{30})=(4.28,1.60)$, in the
energetically relevant asymmetric scission region.

\begin{figure*}[tbp]
  \centering
  \includegraphics[width=0.96\textwidth]{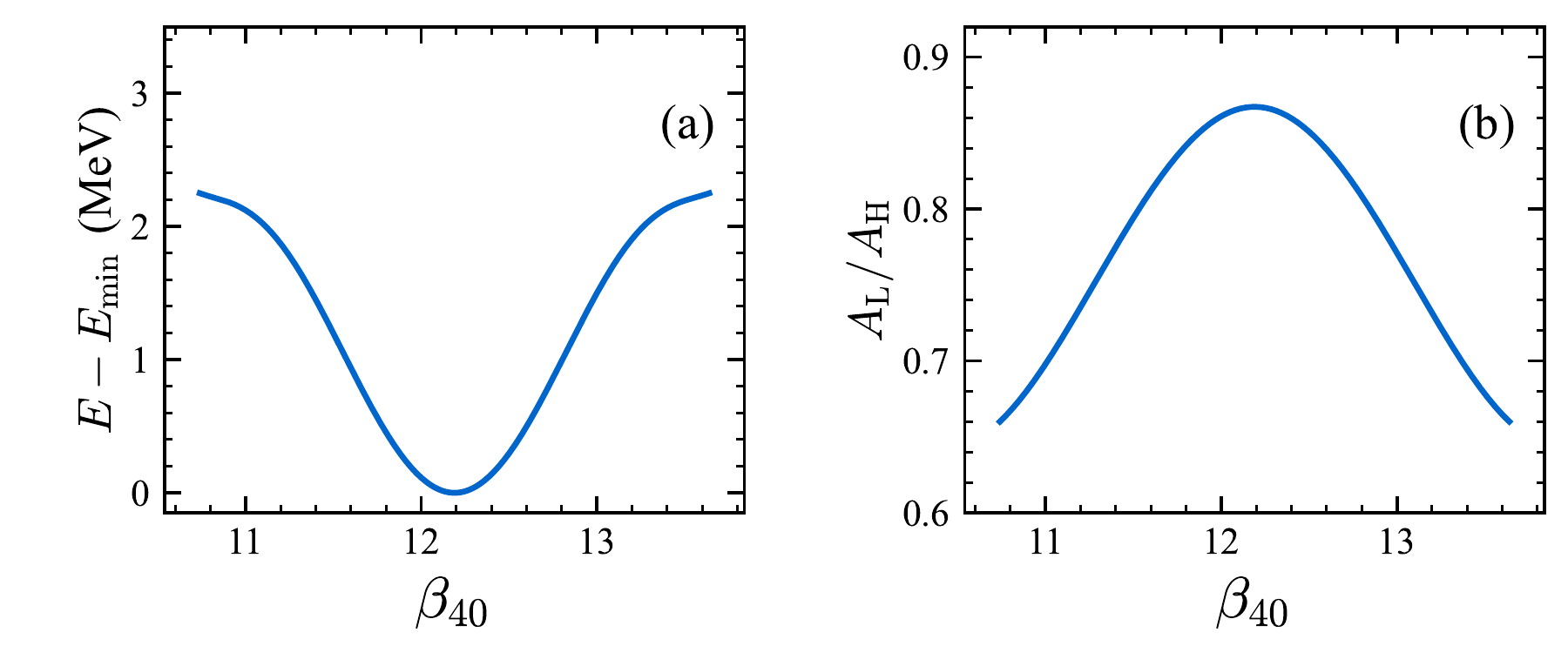}
  \caption{Smoothed illustration of the constrained
  $\beta_{40}$ dependence at $(\beta_{20},\beta_{30})=(4.28,1.60)$:
  (a) energy relative to the minimum, $E-E_{\min}$ and
  (b) light-to-heavy prefragment mass ratio $A_{\mathrm L}/A_{\mathrm H}$.
  The curves are fitted to right-hand scan data and continued
  symmetrically about the sampled minimum at $\beta_{40}=12.19$;
  they do not display all original calculated points.}
  \label{fig:beta40_scan}
\end{figure*}

Figure~\ref{fig:beta40_scan} gives a smoothed illustration of
the energy and light-to-heavy prefragment mass ratio
$A_{\mathrm L}/A_{\mathrm H}$ as functions of $\beta_{40}$.
In the underlying tabulated scan, the lowest-energy sampled solution is
at $\beta_{40}=12.19$. The converged data span
$11.98\lesssim\beta_{40}\lesssim13.76$ and heavy-fragment masses of
approximately 96.4--109.3. At the upper end of this interval the energy
is about 2.33 MeV above the minimum; some intermediate sampled points
lie higher. Here $\beta_{40}$ is the dimensionless mass multipole defined
by Eq.~\eqref{eq:deformation}, normalized with the radius of the whole
$A=180$ system, rather than the deformation of an individual prefragment.

Thus, a given point in the reduced $(\beta_{20},\beta_{30})$ space may
correspond to a range of fragment masses arising from fluctuations in the omitted transverse collective shape degrees of freedom.
The mass obtained from the lowest-energy sampled solution represents the energetically favored
central value, whereas configurations leading to masses farther from
this value are increasingly suppressed as their potential energy rises.
This is consistent with scission-point descriptions in which fragment
deformations beyond the main mass-asymmetry coordinate contribute to
the calculated mass and charge distributions \cite{Pasca2020}.

This behavior motivates replacing the single mass associated with each
$(\beta_{20},\beta_{30})$ configuration by a finite distribution around
the value from the two-dimensional calculation. In the mass-yield calculation, we approximate
the combined effect of the omitted transverse collective fluctuations
by a normalized Gaussian kernel centered at this mass, with phenomenological width
$\sigma_A=4.0$~u. The Gaussian form reflects the dominant weight of
configurations near the energy minimum and the reduced contribution of
configurations farther from it. The adopted width is a phenomenological
estimate guided by the mass variation found in the constrained
$\beta_{40}$ scan, rather than a fitted collective parameter.

\begin{figure*}[tbp]
  \centering
  \includegraphics[width=0.82\textwidth]{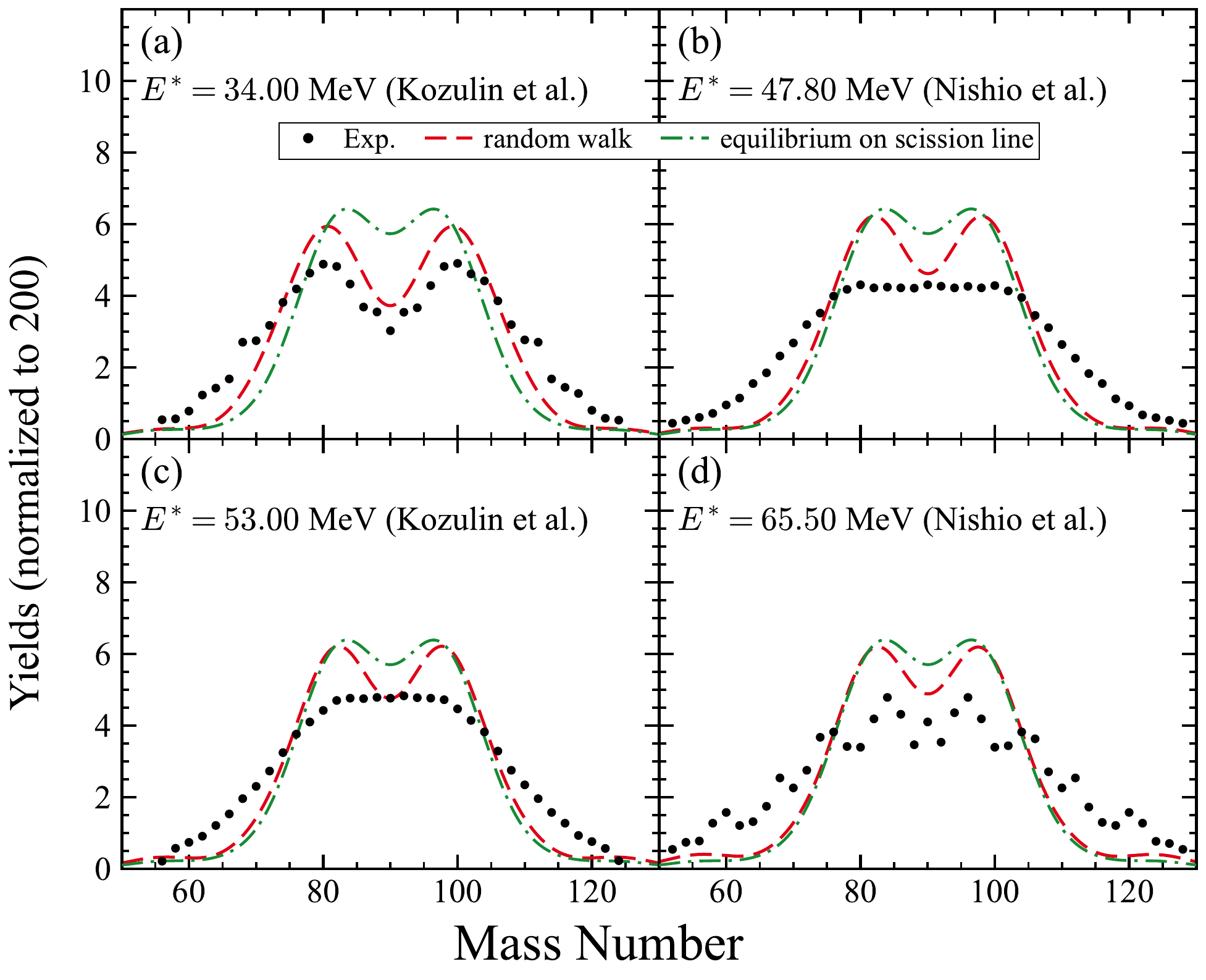}
  \caption{Calculated fission-fragment mass distributions of \Hg{} compared with experimental data
  (black circles) \cite{Nishio2015,Kozulin2022}.
  Panels (a)--(d) show $E^\ast = 34.00$, 47.80, 53.00, and 65.50 MeV, respectively.
  The red dashed curve denotes random-walk (stochastic master-equation) shape evolution,
  and the green dash-dotted curve denotes the equilibrium distribution on the scission line.
  Both calculated distributions include the Gaussian mass folding with $\sigma_A=4.0\,\mathrm{u}$.}
  \label{fig:mass_exp_comparison}
\end{figure*}

With this prescription, the fragment mass yields are obtained by combining the probability
distribution along the scission line with the prefragment masses extracted
from the self-consistent densities. Because the effective barrier top and the geometrical scission line are
located very close to each other in $^{180}$Hg, we also compare the stochastic master-equation result with
the separate equilibrium prescription on the same scission line. The stochastic master-equation calculation uses the
probability distribution shown in Fig.~\ref{fig:scission_probability},
whereas the equilibrium result is obtained from
Eq.~(\ref{eq:equilibriumweight}) on the same set of scission
configurations.

The resulting mass distributions are compared with the available
fusion-fission data in Fig.~\ref{fig:mass_exp_comparison}. The Gaussian
folding broadens the calculated peaks and improves their overall agreement
with the experimental widths while retaining the asymmetric peak positions
and their excitation-energy dependence. Both the stochastic master-equation and equilibrium-on-scission-line calculations predict dominant peaks close to the experimentally observed
$A_{\mathrm L}/A_{\mathrm H}\simeq80/100$ split.
The excitation-energy dependence is visible in the relative peak heights.
With increasing $E^\ast$, the yield around the symmetric division
$A\simeq90$ is enhanced, while the asymmetric peaks are reduced.
This trend reflects the increasing statistical access to configurations
away from the lowest asymmetric portion of the scission line, although
the asymmetric branch still carries the dominant weight in \Hg{}.

Another feature of Fig.~\ref{fig:mass_exp_comparison} is the convergence
of the stochastic master-equation and equilibrium-on-scission-line results
at high excitation energy. This is consistent with the PES analysis above.
For \Hg{}, the effective barrier-top region and the scission line are close
in the $(\beta_{20},\beta_{30})$ landscape, leaving only a limited interval
for collective evolution before scission. Consequently, the fragment-mass
distribution is strongly influenced by the energetic structure of the
scission line itself.
The similarity of the two calculations, particularly at higher excitation
energy, indicates that much of the mass-asymmetry information is already
encoded in the energetic structure of the scission line, while the preceding
stochastic evolution mainly modifies the relative population of nearby
scission configurations.
This differs from the macroscopic-microscopic Brownian shape-motion picture
of Refs.~\cite{Moller2012,MollerRandrup2015}, where the yield is mainly
interpreted in terms of the global potential-energy landscape.
The present result indicates that, for \Hg{}, the fragment mass distribution
is governed not only by the topology of the PES but also by the detailed
energetic structure of the scission boundary.

\section{Summary and Perspective}
\label{sec:summary}

In summary, the present work provides a covariant, density-based
description of the asymmetric fission of $^{180}$Hg, connecting the
microscopic deformation landscape, master-equation shape evolution, and
the properties of the nascent fragments with the observed asymmetric
mass distribution. The potential-energy surface in the
$(\beta_{20},\beta_{30})$ plane exhibits a pronounced
reflection-asymmetric valley. The scission boundary is identified from
the self-consistent density using a geometrical neck criterion, while
the prefragment masses, charges, and intrinsic deformations are
determined independently from the corresponding density distributions.
The energetically favored scission configurations involve strongly
deformed prefragments and lead to masses close to the experimentally
observed $A_{\mathrm L}/A_{\mathrm H}\simeq80/100$ split.

The evolution toward scission is described by a master equation, with
relative transition rates determined from deformation-dependent
microscopic level densities based on the MDC-RHB single-particle spectra
and pairing properties. The resulting stochastic dynamics preserves the
preference for asymmetric scission configurations over the considered
excitation-energy range. An important feature of $^{180}$Hg is the close
proximity of the effective barrier top and the geometrically defined
scission line. 

The preferred asymmetric mass division persists in both the
dynamical and equilibrium prescriptions, identifying the
scission-line structure as an important constraint on the fragment
masses. Within the present model, the preceding shape evolution
mainly modifies the relative populations of scission configurations.

An additional constraint analysis in the $\beta_{40}$ direction shows
that variations of this higher-order shape degree of freedom can produce
an additional spread of the prefragment masses.

Further improvements should include the explicit role of collective
inertias, fluctuations beyond the $(\beta_{20},\beta_{30})$ collective
space, and post-scission neutron emission. These ingredients will be
important for a more quantitative description of the mass-width
systematics and for testing the same mechanism in neighboring
neutron-deficient Hg, Pb, and Pt nuclei.

\begin{acknowledgments}
This work was partly supported by the National Key R\&D Program of China
(Grants No. 2024YFE0109800 and No. 2023YFA1606500),
the National Natural Science Foundation of China
(Grants No. 12447101, No. 12375118, No. 12435008, and No. W2412043),
the Strategic Priority Research Program of the Chinese Academy of Sciences
(Grants No. XDB0920000 and No. XDB1550000),
and the Russian Science Foundation (Grant No. 25-42-00018).
The results described in this paper were obtained on the High-Performance Computing Cluster of ITP-CAS and
the ScGrid of the Supercomputing Center, Computer Network Information Center of Chinese Academy of Sciences.
\end{acknowledgments}

\section*{Data Availability}

The data that support the findings of this article are not publicly available. 
The data are available from the authors upon reasonable request.

\bibliographystyle{apsrev_full_titles}
\bibliography{180Hg_Version_0.9}

\end{document}